# Public Support for Misinformation Interventions Depends On Perceived Fairness, Effectiveness, and Intrusiveness

Catherine King,* Samantha C. Phillips,* and Kathleen M. Carley

**Abstract.** The proliferation of misinformation on social media has concerning possible consequences, such as the degradation of democratic norms. While recent research on countering misinformation has largely focused on analyzing the effectiveness of interventions, the factors associated with public support for these interventions have received little attention. We asked 1,010 American social media users to rate their support for and perceptions of ten misinformation interventions implemented by the government or social media companies. Our results indicate that the perceived fairness of the intervention is the most important factor associated with support, followed by the perceived effectiveness of that intervention and then the intrusiveness. Interventions that supported user agency and transparency, such as labeling content or fact-checking ads, were more popular than those that involved moderating or removing content or accounts. We found some demographic differences in support levels, with Democrats and women supporting interventions more and rating them as more fair, more effective, and less intrusive than Republicans and men do, respectively. It is critical to understand which interventions are supported and how they are perceived, as public opinion can play a key role in the rollout and effectiveness of policies.

## 1 Introduction

There is widespread concern among Americans about the consequences of online misinformation (Mendez 2024). Scholars have linked misinformation to the degradation of democratic institutions and norms (Ecker et al. 2024; Tucker et al. 2018) and the spread of conspiracy theories (De Coninck et al. 2021; Enders et al. 2023; Rottweiler and

*. These authors contributed equally to the work.

*Journal of Online Trust and Safety*, August 2026, page 1 of 48

Gill 2022), although the extent of its harms remains a topic of active debate (Allen and Rand 2024; Budak et al. 2024; Ecker et al. 2024). In response, a growing body of work has examined interventions aimed at curbing misinformation, including social corrections (Badrinathan and Chauchard 2024), warning labels (Mena 2020), accuracy prompts (Pennycook and Rand 2022), and policy-based efforts (Radu 2020). Several review articles synthesize this literature (Blair et al. 2024; Courchesne, Ilhardt, and Shapiro 2021; Helmus and Kepe 2021), typically evaluating interventions based on their ability to reduce the creation, spread, or belief in false content. However, effectiveness alone is insufficient to ensure the success of misinformation interventions in real-world settings. Public trust and willingness to engage with these interventions are critical.

Indeed, researchers and policymakers have emphasized the importance of public participation in designing responses to misinformation (Donovan 2020; Koulolias et al. 2018). Public opinion can play a pivotal role in motivating public policy (Burstein 2003). Furthermore, social media platforms are unlikely to implement unpopular countermeasures, as they are responsive to the desires of their users and any potential revenue implications (Liu, Yildirim, and Zhang 2022). Understanding when and why members of the public support different types of interventions is thus essential to addressing misinformation at scale.

Previous work has identified several factors that shape support for misinformation interventions and efforts to improve the quality and civility of social media discourse more broadly. Personal attributes, such as partisanship, trust in institutions, and previous exposure to misinformation or interventions, play a key role (Appel, Pan, and Roberts 2023; Martin and Hassan 2022; Munzert et al. 2025; Newman et al. 2025; Saltz et al. 2021; Solomon et al. 2024). For example, Democrats tend to be more willing to censor to mitigate potential harms (Appel, Pan, and Roberts 2023; Munzert et al. 2025; Saltz et al. 2021; Solomon et al. 2024). Context also shapes support. In particular, content that is tied to obvious harms (e.g., calls for violence) or targets vulnerable groups elicits greater support for intervention (King and Carley 2025; Kozyreva et al. 2023; Pradel et al. 2024; Munzert et al. 2025). Furthermore, there is variation in support across interventions. People generally prefer harm-based, content-focused, reversible measures over account-focused, permanent consequences (Kozyreva et al. 2023; Kubin, Sikorski, and Gray 2025; Solomon et al. 2024). Yet there has been little consideration of how perceived qualities of misinformation countermeasures might predict preferences.

In this study, we investigate three features—fairness, effectiveness, and intrusiveness—of interventions that have previously been identified as key predictors of public policy support across various domains (Grelle and Hofmann 2024), including climate change initiatives (Bergquist et al. 2022; Huber, Wicki, and Bernauer 2020) and public health interventions (Bos et al. 2015; Diepeveen et al. 2013). Perceived fairness is critical in predicting support for misinformation interventions, as content moderation efforts often raise concerns about potential bias or disproportionate impacts on specific groups (Chuai et al. 2024; Donovan 2020; Mosleh et al. 2024; Radu 2020; Rich, Milden, and Wagner 2020; Vogels 2022). Effectiveness is important because support is likely to depend on whether the intervention is seen as capable of meaningfully reducing misinformation exposure or harm (Tay et al. 2023; Tsang and Zhou 2025). Finally, intrusiveness matters because interventions perceived as overreaching or invasive are less likely to be supported (Khatiwada et al. 2025; Saltz, Leibowicz, and Wardle 2021).

Furthermore, the entity responsible for implementing an intervention may shape how individuals weigh concerns about fairness, effectiveness, and intrusiveness. For instance, perceptions of whether an intervention infringes on free speech can depend on the implementing actor, making fairness particularly salient. In the United States, citizens have expressed greater concern about government restrictions on speech than similar actions taken by private companies (Mitchell and Walker 2021). A review article on public policy acceptance across various domains finds that the desire for government support positively interacts with favorable factors like perceived effectiveness and fairness, and helps reduce the negative impact of factors such as perceived intrusiveness, thus influencing overall policy support (Grelle and Hofmann 2024).

Therefore, our first research question focuses on how perceived fairness, intrusiveness, and effectiveness of an intervention are associated with the support for a given intervention, depending on whether the measure is implemented by social media platforms or the government.

**RQ1.1** To what extent is a misinformation intervention's perceived fairness, intrusiveness, and effectiveness associated with support for that intervention?
**RQ1.2** How do the attributes people consider when forming preferences change due to the implementer of the intervention?

Next, we compare the general support, perceived fairness, perceived effectiveness, and perceived intrusiveness for each intervention. This allows us to identify which strategies are viewed more or less favorably on average, and where opinion is most divided and influenced by the implementation institution.

**RQ2** What is the average and variance in support, perceived fairness, perceived intrusiveness, and perceived effectiveness for each intervention?

Furthermore, certain segments of the US population on social media may be more or less accepting of misinformation interventions. Some demographic groups might want more government oversight or have a better understanding of the misinformation issue, which can influence public policy support levels (Grelle and Hofmann 2024). Understanding demographic and partisan differences in intervention support informs public messaging and intervention design, as well as reveals larger trends in the values involved in policy support judgments. Therefore, we ask,

**RQ3.1** How strongly are demographic differences associated with support for misinfor-mation interventions?
**RQ3.2** Is support correlated with different attributes for different demographic groups?

To address these research questions, we surveyed 1,010 US residents who use social media at least once a week. These participants represent those most likely to have encountered and be impacted by online misinformation and interventions (Meshi and Molina 2025). We randomized whether participants were told the government or social media platforms would implement the interventions. Participants rated their support for and perceived effectiveness, fairness, and intrusiveness of each intervention. We included ten interventions designed to encompass some of the most studied measures in the literature.

This research contributes to current debates about the feasibility and design of misinformation interventions by providing systematic evidence on how the public evaluates trade-offs among competing priorities. In doing so, it offers insight into the conditions under which interventions are likely to be perceived as legitimate or acceptable, and how these perceptions may differ across contexts and communities. To our knowledge, this is the first study to compare public support for a broad array of intervention types implemented by both government entities and platforms. In addition, we develop a typology of misinformation interventions to assist in comparing and contrasting support for various policies and implementers.

## 2 Intervention Typology

In this study, we examined ten interventions that could be implemented by either a social media platform or a government entity. These interventions were selected to represent a broad range of possible countermeasures. To develop this list of comprehensive interventions, we reviewed the existing literature on previous categorizations of interventions and the life cycle of misinformation on social media to better understand when and where interventions should be deployed (King 2025).

### 2.1 Previous categorizations

While there are multiple intervention categorizations proposed in previous review articles, there is no common typology (Blair et al. 2024; Courchesne, Ilhardt, and Shapiro 2021; Gwiaździński et al. 2023; Helmus and Kepe 2021; Kozyreva et al. 2024). Some review articles, such as Courchesne et al. (2021), categorize only interventions that have been publicly announced as implemented by various social media platforms. These platform-only intervention categories include advertising policy, content/account moderation, content labeling, content reporting, and content distribution (Courchesne, Ilhardt, and Shapiro 2021). Other prominent review articles categorize interventions not only in terms of which parts of the platforms are affected but also by what part of the social media misinformation pipeline is targeted (such as the creation, spread, or belief in misinformation) (Blair et al. 2024; Kozyreva et al. 2024). For example, Blair et al. (2024) categorize 11 types of interventions into four main groups: *Institutional*, which targets the original creators and distributors of misinformation by altering the platforms, training journalists, etc.; *Sociopsychological*, which targets the spread of misinformation by discouraging users from sharing it; *Informational*, which targets the belief in misinformation through prebunking, debunking, or tagging content; and *Educational*, which aims to prevent belief in misinformation (Blair et al. 2024). Similarly, Kozyreva et al. (2024) classify nine types of interventions into three main categories: *Nudges*, which target the spread of misinformation by discouraging users from sharing it; *Refutation Strategies*, which target the belief in misinformation through fact-checking, debunking, or tagging content; and *Boosts and Educational Interventions*, which aim to prevent belief in misinformation (Kozyreva et al. 2024).

Many of these previously defined categorizations overlap in several areas. However, some potential intervention categories are absent in many review papers, such as user-led interventions like social corrections or user reporting (Badrinathan and Chauchard 2024; King, Phillips, and Carley 2025) and external structural measures like data sharing and transparency (Helmus and Kepe 2021). A recent bibliometric analysis of the literature on misinformation interventions combined the categorizations of several prominent review articles and added a separate category for both user-based measures and external, institutional measures (King, Carragher, and Carley 2025). To effectively categorize interventions, it is crucial to consider all possible individuals or organizations involved (users, platforms, governments, or other institutions) in addition to the specific part of the misinformation life cycle being targeted and how.

### 2.2 Targeting the misinformation pipeline

The life cycle of misinformation content on social media is often referred to as the "misinformation pipeline" (Ciampaglia 2018). Previous definitions tend to include the creation and dissemination of a misinformation message, in addition to what happens after the message has spread (Ciampaglia 2018; Ng and Taeihagh 2021; Wardle and Derakhshan 2017). We define the misinformation pipeline as consisting of three main phases: **creation**, **spread**, and **belief**. This pipeline informs the development of targeted intervention categories for each stage of the misinformation life cycle.

#### 2.2.1 Creation

This phase typically refers to the original inception of the misinformation message (Wardle and Derakhshan 2017) and the accounts that will share it (Ng and Taeihagh 2021). There are two related components of this stage: the *network creation* and the *content creation*. Network creation refers to the creation of accounts that will initially disseminate the message, along with the networks they form to further spread the message (Ng and Taeihagh 2021). Malicious users may obtain or hijack existing accounts, or create a set of coordinating bot accounts (Ng and Taeihagh 2021). Content creation is the process of developing the original misinformation message and transforming it into a media product (Ng and Taeihagh 2021; Wardle and Derakhshan 2017).

Potential interventions in the creation step of the misinformation pipeline typically focus on detecting fake accounts and inauthentic activity (Ng and Taeihagh 2021), or on the algorithmic detection of misinformation content (Ciampaglia 2018).

#### 2.2.2 Spread

This phase refers to how platforms and users distribute misinformation content (Ciampaglia 2018; Wardle and Derakhshan 2017). There are two related components: *sharing* and *amplification*. The initial sharing refers to the direct act of sharing the misinformation content with others, such as by posting the content, messaging the content directly to specific users, or forwarding the message. Further amplification refers to how engagement with content or algorithmic bias can further spread the message (Ciampaglia 2018). Malicious users can also coordinate with other users or use bot accounts to artificially boost engagement with a post to further its initial spread (Ng and Taeihagh 2021).

Potential interventions in this step of the misinformation pipeline typically concentrate on introducing friction before regular users share content without thinking (Kozyreva et al. 2024) or implementing algorithmic downranking and other platform alterations to reduce artificial amplification (Wardle and Derakhshan 2017).

#### 2.2.3 Belief

This phase refers to the false beliefs that may arise from the spread of misinformation (Ciampaglia 2018). Belief has two related components: *verification* and *prevention*. Verification refers to the verification of the posted and spread content. Verification can happen through automated systems or human fact-checkers (Ciampaglia 2018). Fact-checking can lead to content labels, warnings, or removal. Prevention interventions include inoculating or warning users about specific misinformation content or techniques they may encounter (Lewandowsky and van der Linden 2021), as well as educational efforts aimed at improving media literacy skills and overall competencies (Altay, De Angelis, and Hoes 2024; Jeong, Cho, and Hwang 2012).

Potential interventions in this step of the misinformation pipeline typically focus on correcting, labeling, or removing false content (Walter and Murphy 2018), educating and warning the public about the misinformation they may encounter (Lewandowsky and van der Linden 2021), and promoting trust in reliable news sources (Altay, De Angelis, and Hoes 2024).

### 2.3 Proposed intervention categorization

After reviewing the literature on previous categorizations and the misinformation pipeline, we developed six general categories of countermeasures, as shown in Table 1. These categories are primarily classified by the part of the misinformation life cycle targeted and the affected environments.

Table 1: Intervention categories grouped by the targeted phase of the misinformation pipeline.

| Category | Target | Definition |
|---|---|---|
| Account Moderation | Creation | The moderation of user accounts such as by suspending, banning, or limiting users |
| Content Moderation | Spread | The moderation of content such as by removing, downranking, or debunking content |
| Content Distribution | Spread | Efforts affecting the distribution or sharing of content such as by using redirection, accuracy prompts, or friction |
| Content Labeling | Belief | The use of labeling or misinformation disclosure to notify users, provide additional context, or fact-check |
| Media Literacy | Belief | Training efforts aimed at improving the public's media literacy and critical thinking skills |
| External Structural Responses | Various | Measures taken outside the platform ecosystems such as regulation, data sharing, and investing in journalism |

The first four general categories focus on interventions that occur on platforms, and these can be initiated by the social media companies themselves, or governments can exercise oversight through regulation or other legislative measures. Platform interventions often involve algorithmic changes regarding what accounts and content can be created or distributed on the platform (account moderation, content moderation, content distribution) or front-end design changes concerning how content is displayed or can be interacted with after it has spread (content labeling). More specifically, account moderation strategies typically target the creation phase of the misinformation pipeline by controlling account creation and determining which accounts are permitted to post content. Content moderation interventions usually target the spread of misinformation by reducing or removing algorithmic amplification, while content distribution interventions focus on managing the sharing of misinformation by user accounts. Finally, content labeling interventions occur after misinformation has already been shared, and they employ verification techniques to correct or warn about false or misleading information.

The last two categories, media literacy and external structural responses, occur off-platform and can be implemented by platforms, governments, or various civic organizations. Media literacy and other educational initiatives primarily focus on preventing the belief phase in the misinformation pipeline, while structural strategies may target any part of the pipeline. These external structural responses are measures taken outside platform ecosystems to support a healthier information environment.

We identified 1–2 representative interventions per category from the literature to present to study participants. Participants were told in advance whether the implementer of the intervention was social media platforms or government entities, explicitly mentioning that the determiner of what misinformation is would fall on the intervention implementer (e.g., platforms could fact-check internally or use an external, independent organization). The interventions are described in Table 2.

Table 2: Selected misinformation interventions. Text in [brackets] was included when the specified implementer was the government.

| Category | Intervention | Reference(s) |
|---|---|---|
| Account Moderation | 1. [Require social media companies to] permanently ban users who post misinformation a certain number of times. | (Gao and Thebault-Spieker 2024; Rauchfleisch and Kaiser 2024) |
| Content Moderation | 2. [Require social media companies to] remove posts verified to contain misinformation. | (Jiang et al. 2023) |
| | 3. [Require social media companies to] de-emphasize posts that are verified to contain misinformation. | (Gillespie 2022) |
| Content Distribution | 4. [Require social media companies to] temporarily delay users posting content the user did not open or spent less than a certain amount of time viewing. | (Fazio 2020; Porterfield 2020) |
| | 5. [Require social media companies to] put all advertising through a fact-checking process. | (Helmus and Kepe 2021) |
| Content Labeling | 6. [Require social media companies to] notify users if they posted content verified to contain misinformation. | (Courchesne, Ilhardt, and Shapiro 2021) |
| | 7. [Require social media companies to] publicly label posts verified to contain misinformation with information about and from verified sources. | (Papakyriakopoulos and Goodman 2022; Yadav 2021) |
| Media Literacy | 8. Invest in digital media literacy and promote educational content about detecting misinformation on and offline. | (Guess et al. 2020; Roozenbeek et al. 2022) |
| External Structural Responses | 9. Promote and invest in local media, which is thought to be most in tune with local norms, culture, and context. | (Bradshaw and Neudert 2021; Toff and Mathews 2024) |
| | 10. [Require social media companies to] regularly release data and/or internal research reports about misinformation prevalence, spread, and mitigation to the public and outside researchers. | (Pasquetto et al. 2020) |

## 3 Data and Methods

Our sample (convenience) was recruited from Amazon Mechanical Turk via the MTurk Toolkit on CloudResearch, which draws from pre-screened, CloudResearch-approved MTurk workers to ensure quality responses.[1] Each participant was randomly assigned

1. https://www.cloudresearch.com/products/turkprime-mturk-toolkit/

to see interventions implemented by either the government or social media companies. Each participant saw a random subset of 8 (of 10) interventions. The responses analyzed in this paper were part of a broader survey on misinformation interventions, and we only report the relevant section as specified in our pre-registration.[2] Part 1 of the survey focused on misinformation exposure and the likelihood of engaging in countering behaviors (King, Phillips, and Carley 2025), while the responses from the second half of the survey are reported in this paper.

There were 1,684 participants who started the survey, 661 of whom were removed before any responses to outcome measures were recorded. Of the 661 participants removed, 504 failed Qualtrics' duplicate response or bot detection, 67 failed to provide consent or indicate they are 18 years old or older, 79 failed a screening question (participants were asked to select their favorite season and then indicate which holiday falls in the indicated season), and 6 passed the screening but quit before responding further. Only 5 participants were removed because they did not fulfill the American residency or social media user requirements. Of the 1,023 participants who qualified for the survey and responded to at least some outcome measures, 13 quit before completing the survey.

### 3.1 Demographic Variables

We collected standard demographic variables including age, gender, race, education, income, political party affiliation, and political ideology. We additionally collected self-reported exposure to misinformation and frequency of platform usage for each platform. Our final sample contained 465 females, 520 males, and 25 people who identified as transgender, non-binary, or preferred not to self-identify. We had 28 participants aged 18–24, 225 aged 25–34, 338 aged 35–44, 186 aged 45–54, 148 aged 55–64, and 85 aged 65 or older. Finally, of the 1,010 participants in our final sample, 471 identified as Democratic, 236 identified as Republican, and 303 identified as Independent/Other. See Appendix E, Table 15 for all sample statistics.

### 3.2 Measures

*Support for intervention(s).* We asked participants to rate a subset of interventions as strongly support, somewhat support, neither support nor oppose, somewhat oppose, or strongly oppose. These responses are coded from 1 to 5 (least to most support).

*Perceived fairness of intervention(s).* We asked participants to rate a subset of interventions as very fair, somewhat fair, neither fair nor unfair, somewhat unfair, or very unfair. These responses are coded from 1 to 5 (least to most fair).

*Perceived intrusiveness of intervention(s).* We asked participants to rate a subset of interventions as very intrusive, somewhat intrusive, neither intrusive nor unintrusive,

2. https://osf.io/b2yjt/

somewhat unintrusive, or very unintrusive. These responses are coded from 1 to 5 (least to most intrusive).

*Perceived effectiveness of intervention(s).* We asked participants to rate a subset of interventions as very effective, somewhat effective, neither effective nor ineffective, somewhat ineffective, or very ineffective. These responses are coded from 1 to 5 (least to most effective).

## 3.3 Analyses

### 3.3.1 RQ1

We indicated in our pre-registration that we planned to run a multilevel model to account for random effects (i.e., random slope and intercept) of interventions and participants, as each participant saw 8 of the 10 selected interventions, drawn randomly, and multiple participants rated each intervention.

We also pre-registered that if this model did not converge, we would fit an OLS regression model with robust standard errors clustered on participants and interventions. Since the multilevel model did not converge, the model output reported in this article is an OLS regression with robust standard errors clustered on participants and interventions (see Appendix A, Table 3). Because the model includes interactions, coefficients are interpreted as conditional associations; we summarize overall associations using average marginal effects calculated using the "marginaleffects" R package (Arel-Bundock, Greifer, and Heiss 2024). We additionally ran planned robustness checks by including participants who responded to a part of the survey but did not complete it (see Appendix A, Table 5). This does not change the direction or significance of the effects found in the primary model.

We calculated adjusted fractional Bayes factors with Gaussian approximations for the primary models using the BFPack R package (Mulder et al. 2021). We report BF10 for each estimate where the alternative hypothesis is directional based on the sign of the estimate (i.e., $b < 0$, $b > 0$) and the null hypothesis is $b = 0$. Thus, if $BF > 1$, the evidence is more consistent with the alternative hypothesis; if $BF < 1$, the evidence is more consistent with the null hypothesis.

### 3.3.2 RQ2

We indicated in the pre-registration that we would conduct descriptive analyses of the average and spread of the ratings for each intervention. In addition, we conducted ad hoc t-tests (see Appendix B, Table 7) comparing ratings of each intervention if they were implemented by platforms or by governments (significant at $p = 0.0011$ with Bonferroni correction for 44 t-tests total).

#### 3.3.3 RQ3.1

For our analysis of individual differences in support and perceptions of interventions, we ran the planned participant regression models as specified in our pre-registration. In Appendix C, Table 8 shows the main results, and Tables 9 and 10 show the same models run with different codings of the partisanship variable. In addition, to complement the pre-registered regressions, we ran one-way ANOVA tests comparing average support, perceived fairness, perceived effectiveness, and perceived intrusiveness across categories for each demographic variable measured categorically (i.e., partisanship, gender, age, income, education, ideology, misinformation exposure frequency, and number of platforms used weekly) (see Appendix C, Table 11).

#### 3.3.4 RQ3.2: Ad-hoc analysis

Finally, we included partisanship, gender, and number of platforms used weekly interacting with implementer and perceptions of fairness, effectiveness, and intrusiveness to predict support (i.e., added demographic variables to the model used in RQ1) (see Appendix D, Table 13).We also ran the same model with gender, ideology, and number of platforms used weekly included (see Appendix D, Table 14).

## 4 Results

### 4.1 Perceived fairness, effectiveness, intrusiveness, and implementer are associated with support for interventions.

Figure 1 shows the regression coefficients for RQs 1.1 and 1.2. Full regression results and average marginal effects (AMEs) are in Appendix A, Table 3. Perceived fairness is the most strongly associated with support ($AME = 0.584, SE = 0.012, p < 0.001$), followed by perceived effectiveness ($AME = 0.341, SE = 0.011, p < 0.001$) and intrusiveness ($AME = -0.083, SE = 0.007, p < 0.001$).

The average difference in predicted support between the government and the platforms as the implementers is small and not statistically distinguishable from zero ($AME = -0.018$, $SE = 0.015$, $p = 0.23$). However, the relationships between perceived features and support differ across implementers: Fairness is less associated with support when the implementer is government than when the implementer is social media platforms ($\beta = -0.080, SE = 0.024, p < 0.001$), while intrusiveness ($\beta = -0.036, SE = 0.015, p = 0.015$) and effectiveness ($\beta = 0.077, SE = 0.022, p < 0.001$) are more strongly associated.

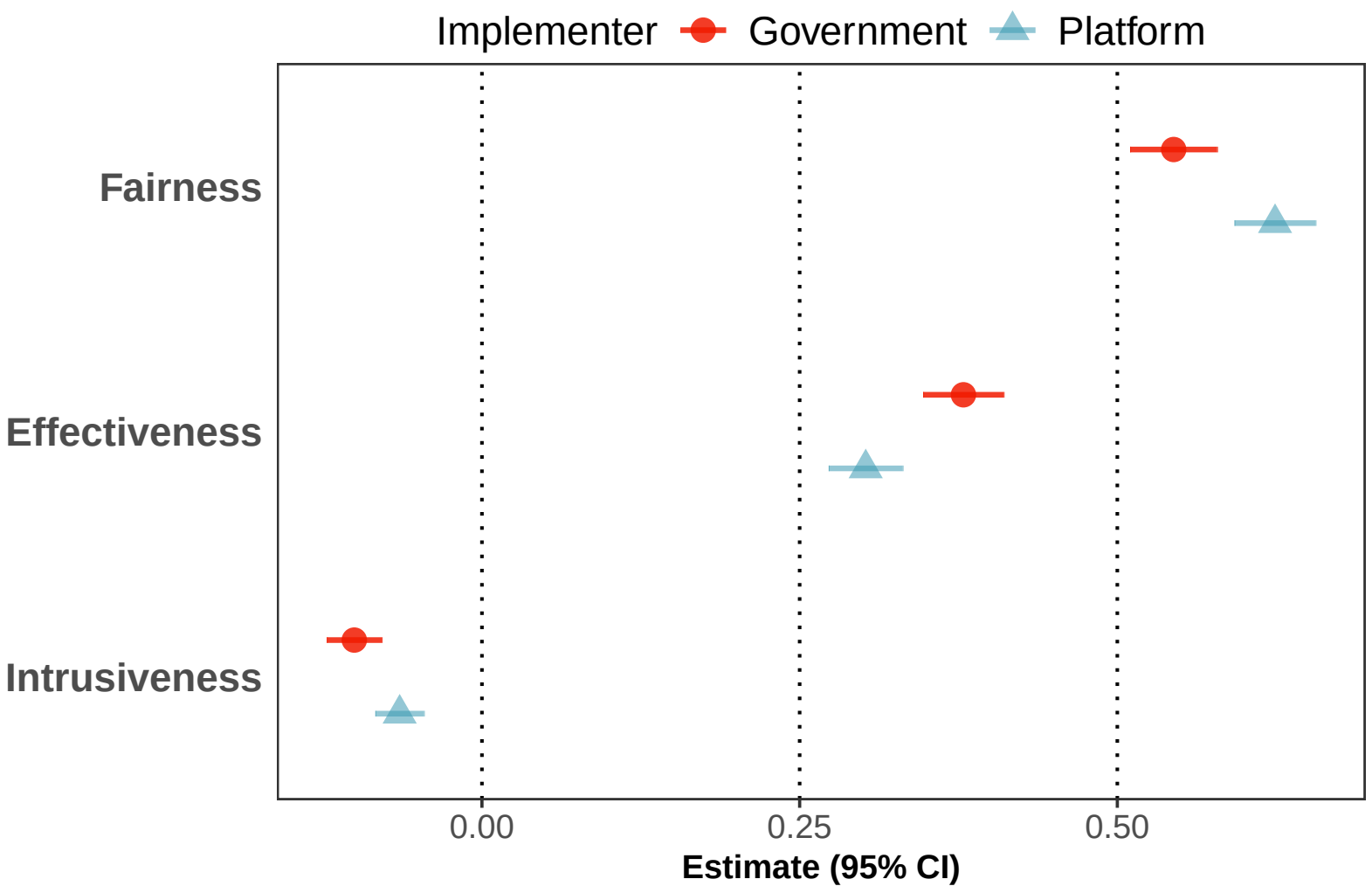

Figure 1: Estimate and 95% CI of the effect of perceived fairness, effectiveness, and intrusiveness on support depending on implementer.

## 4.2 Overall support and perceptions of interventions.

Figure 2 contains the estimate and 95% CI for support, perceived fairness, perceived effectiveness, and perceived intrusiveness for each intervention by each implementer (government and social media platforms). Participants were more supportive of interventions in the content labeling category and less supportive of those in the content distribution or moderation categories.

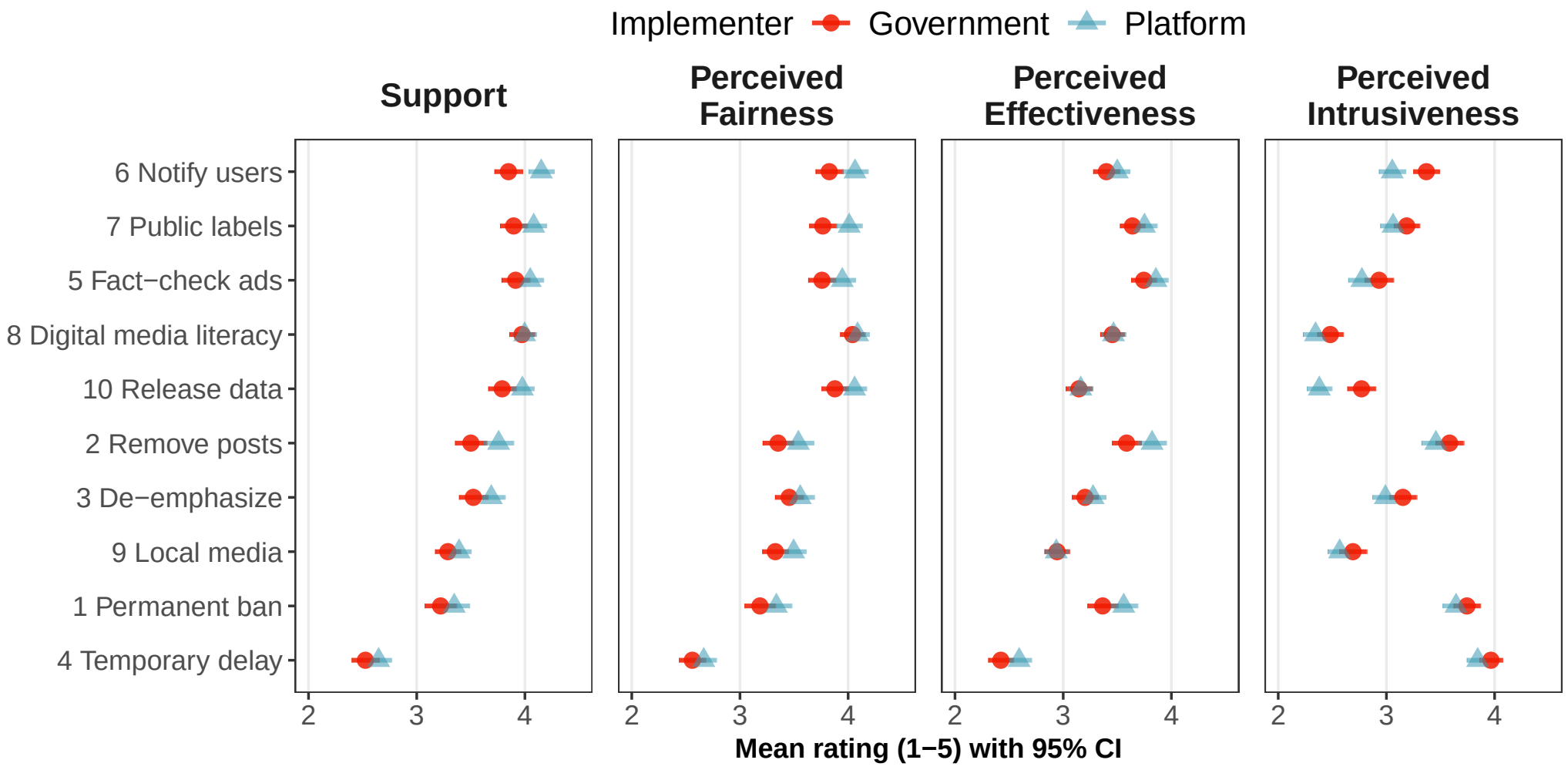

Figure 2: Average and 95% CI support, perceived fairness, perceived effectiveness, and perceived intrusiveness for each intervention (1–10) and implementer (government and platform) on a 1–5 Likert scale.

When comparing the ratings of interventions when implemented by platforms versus governments, we found that overall, people support interventions more ($t$ = 5.48, $p < 0.0001$) and perceive them as more fair ($t = 5.58$, $p < 0.0001$), more effective ($t = 3.63$, $p < 0.001$), and less intrusive ($t = -6.08$, $p < 0.0001$) when implemented by platforms compared to governments. There is not a significant difference in perceptions depending on the implementing entity for most interventions examined separately, with the following exceptions: First, notifying users was supported more ($t = 3.41$, $p < 0.001$) and perceived as less intrusive ($t = -3.59$, $p < 0.001$) if implemented by platforms compared to governments. Second, releasing data and/or internal research is perceived as significantly more intrusive if implemented by governments rather than by platforms directly ($t = -4.4$, $p < 0.0001$). Full ad-hoc t-test results are in Appendix B, Table 7.

### 4.3 Individual differences in support and perceptions of interventions.

Next, we investigate individual differences in support and perceived attributes of misinformation interventions (full regression outputs are in Appendix C, Table 8).Unsurprisingly, we find Democrats support interventions more than Independents/other ($\beta = -0.244, SE = 0.074, p < 0.001$) and Republicans ($\beta = -0.275, SE = 0.106, p < 0.01$) do. We also find Independents/other and Republicans perceive interventions as less fair and effective than Democrats do ($p < 0.05$ for all), while only Independents/other perceive interventions as more intrusive than Democrats do ($\beta = 0.203, SE = 0.066, p < 0.01$). These results are robust to separating the “Independent” and “Other/unaffiliated” categories and mapping partisan categories to corresponding numeric values (compare Appendix C, Tables 9 and 10). Similarly, liberal-leaning participants tend to support interventions more compared to conservative-leaning participants ($\beta = -0.296, SE = 0.037, p < 0.001$). Liberal ideology is also associated with perceiving interventions as more fair, more effective, and less intrusive ($p < 0.001$ for all).

We find men are less supportive of interventions than women are on average ($\beta = -0.185, SE = 0.053, p < 0.001$). Men perceive countermeasures as less fair, less effective, and more intrusive than women do on average as well ($p < 0.05$ for all). The number of platforms visited at least once weekly is positively associated with higher support, perceived fairness, and perceived effectiveness ratings and negatively associated with perceived intrusiveness ratings ($p < 0.001$ for all). On the other hand, higher levels of self-reported misinformation exposure are not associated with anything except lower perceived average effectiveness ratings ($\beta = -0.063, SE = 0.024, p < 0.05$). Moreover, people with higher incomes tend to perceive interventions as slightly more fair ($\beta = 0.039, SE = 0.016, p < 0.05$). Finally, compared to younger participants, older participants are more likely to support interventions ($\beta = 0.045, SE = 0.021, p < 0.05$) and find them fairer ($\beta = 0.067, SE = 0.022, p < 0.01$). Education level is not associated with support level or any perceptions of interventions.

In our exploratory ANOVA analysis to investigate these results further (see Appendix C, Table 11), we found statistically significant differences in support for gender, partisanship, ideology, education, and number of platforms visited at least weekly ($p < 0.05$ for all). Age, income, and misinformation exposure groups do not differ in support. Figure 3 shows the average support level, perceived fairness, perceived effectiveness, and perceived intrusiveness broken down by partisanship, gender, and number of platforms used weekly. Notably, gender, partisanship, ideology, and weekly platform usage are the only variables that differ across all outcomes in both regression and ANOVA analyses (except the ANOVA for gender and intrusiveness).[3] Analogous figures for all of the remaining demographic variables can be found in Appendix C, Figure 5.

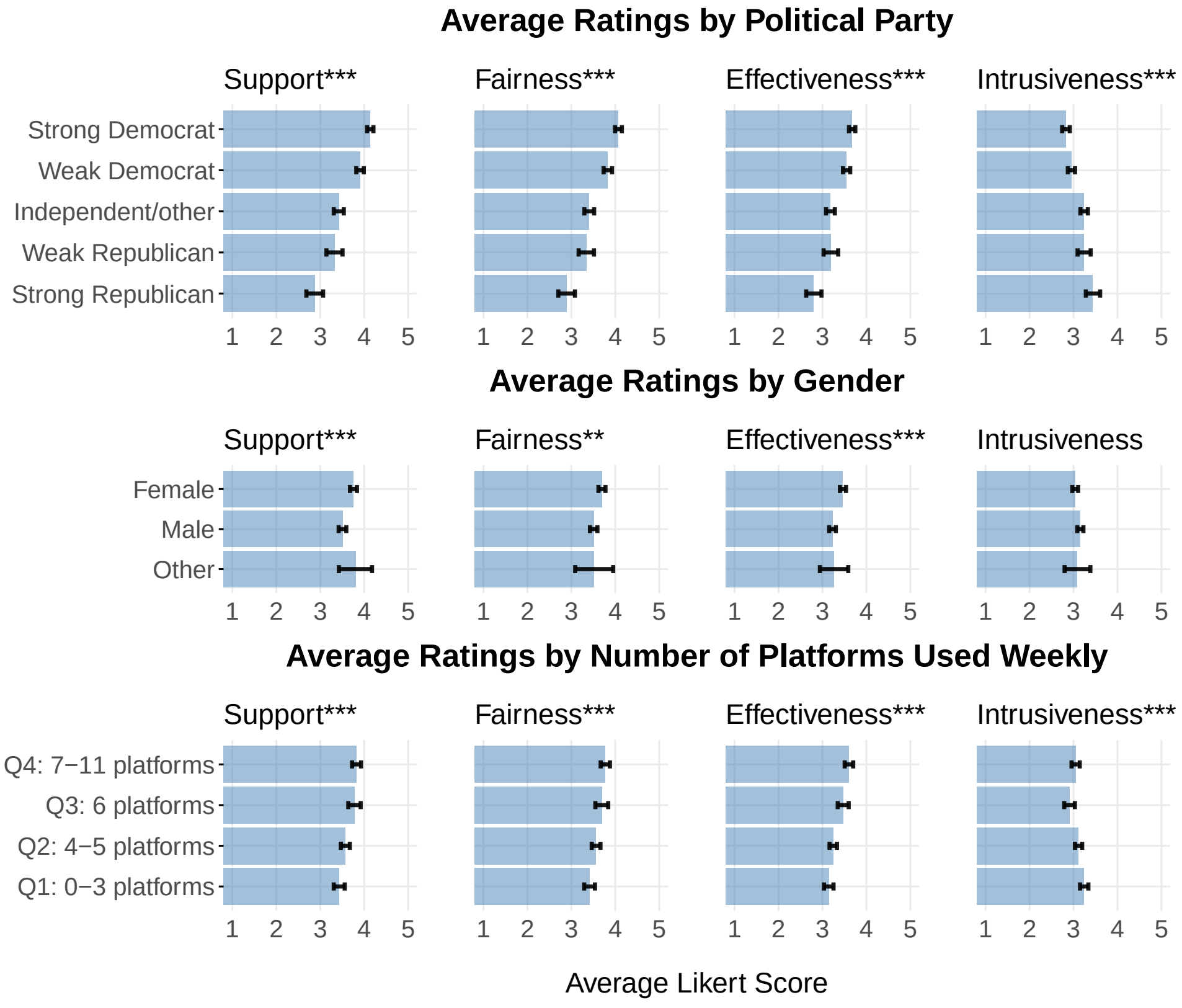

Figure 3: Average ratings (95% CI) by political party, gender, and number of platforms visited weekly broken up into quartiles (Q1: 0–3 platforms, Q2: 4–5 platforms, Q3: 6 platforms, and Q4: 7–11 platforms). One-way ANOVA tests were run on each grouping, with stars indicating the level of significance: $^*p < 0.05$, $^{**}p < 0.01$, and $^{***}p < 0.001$.

**Ad hoc analysis: Partisanship, gender, and platform usage influence features that predict support.**

From our analysis of individual differences in support, we identified partisanship, gender, and number of platforms visited weekly to examine further. We performed an ad hoc analysis to examine how these variables interact with the implementer of the intervention and perceived attributes to predict support (see Appendix D, Table 13). We find that partisanship interacts with implementer and fairness: Republicans care more about fairness than Democrats do ($\beta = 0.065, SE = 0.029, p = 0.026$). Men also were found to

3. We exclude ideology from Figure 3 for conciseness, as partisanship and political ideology are strongly associated; see Appendix C, Table 12.

care more about fairness compared to women ($\beta = 0.048, SE = 0.024, p = 0.043$).

There is also a larger difference in support for interventions implemented by government versus social media companies for Republicans and Independents than there is for Democrats ($\beta = -0.094, SE = 0.039, p = 0.017$; $\beta = -0.103, SE = 0.037, p = 0.005$); Republicans support interventions implemented by governments less. The number of platforms used weekly interacts positively with perceived fairness and intrusiveness ($\beta = 0.011, SE = 0.005, p = 0.040$; $\beta = 0.009, SE = 0.003, p = 0.010$) and negatively with perceived effectiveness ($\beta = -0.011, SE = 0.005, p = 0.036$). This suggests that heavy platform users may value fairness in their own user experience over the potential effectiveness of the interventions when considering support, and they are less deterred by intervention intrusiveness despite being the most active users (see Appendix D, Table 13).

In addition, we ran the same model with political ideology included instead of partisanship (see Appendix D, Table 14). We again find a significant interaction between political ideology and perceived fairness, where conservative-leaning participants weigh fairness more compared to liberal-leaning participants ($\beta = 0.030, SE = 0.010, p = 0.004$).

## 5 Discussion

In this work, we surveyed American social media users to examine public acceptance of interventions against misinformation implemented by the government and social media companies. We found that an intervention's perceived fairness was most strongly associated with support or opposition, followed by perceived effectiveness, and finally perceived intrusiveness. Fairness was a greater concern when the implementer was social media companies rather than the government, while effectiveness and intrusiveness were more salient when the government was the implementer (Figure 1). However, in general, the same intervention implemented by social media companies received more support, was perceived as fairer and more effective, and was viewed as less intrusive than when implemented by the government (Figure 2). These findings may reflect public attitudes toward businesses and government, where companies are more trusted to address misinformation in a timely manner at scale. Alternatively, people may believe that social media companies have a greater responsibility to address misinformation than the government. Platform interventions are also more visible to social media users, whereas government regulation in this domain is largely absent in the US. Additionally, participants who reported heavy platform usage (visiting several platforms per week) also tended to support interventions more and perceive them as both more fair and more effective. The finding that heavy platform users, those who are most likely to encounter interventions, support interventions more than others is promising. More experience with platform interventions, indirectly approximated by our weekly platform usage variable, may increase familiarity and, therefore, positive impressions of interventions (e.g., mere exposure effect (Horowitz et al. 2024; Zajonc 1968)).

Our results further indicate that people desire agency and transparency in misinformation interventions, echoing findings from Saltz et al. (2021) and research from other policy contexts (de Fine Licht 2014; Grelle and Hofmann 2024). They support interventions that provide information to users that they can use when deciding how to interact with certain content, such as notifying them if they have posted misinformation, adding public labels to content containing misinformation, implementing digital media literacy programs, and requiring platforms to release data or internal research reports related to misinformation. There was also strong support for holding advertising accountable through fact-checking. People were generally less supportive of interventions that involve removing or de-emphasizing posts identified as containing misinformation and banning users who repeatedly post misinformation. While banning users is largely not supported, many believed that it would be relatively effective. Belief in effectiveness is simply not enough to support certain interventions that are considered unfair or intrusive. When choosing which interventions to implement, the most supported categories target the **belief** phase of the misinformation pipeline rather than the **creation** of networks or content and their **spread**. These results are consistent with the literature in other policy areas, which finds that the public generally prefers informational interventions over more restrictive measures that target earlier stages in the misinformation pipeline, even though they are often less effective (Diepeveen et al. 2013; Hagmann, Siegrist, and Hartmann 2018).

Interestingly, two of the least supported interventions do not directly involve any censoring activity. Promoting and investing in local media was perceived to be largely ineffective. It may be that this intervention was too vague for participants to envision how it could help mitigate misinformation. Finally, the least popular intervention by a large margin was temporarily delaying users when they attempt to post content they did not view or had barely viewed. It was rated as the least fair and effective and the most intrusive. This result is fairly unexpected considering that some platforms currently implement this intervention, including Facebook and X (formerly Twitter) (Clark 2021; Porterfield 2020). Social media affords instantaneous communication and content, which could drive impatience for even the slightest inconvenience or delay (e.g., commercial breaks (Bruun 2019)). Therefore, when employing nudge-based approaches like accuracy prompts (Pennycook and Rand 2022) or temporary delays in posting, it is imperative to minimize intrusiveness in the user experience, such as through design choices.

Average support for and perceptions of interventions did not significantly differ depending on implementer for most of the interventions (when tested separately), with a few key exceptions. First, people were more likely to support notifications to users who post misinformation when implemented by platforms rather than when implemented by the government (i.e., required of platforms through government regulation). In addition, participants rated this intervention as significantly more intrusive if implemented by the government compared to platforms. It may be that the lack of clarity about the government's role in notifying users of misinformation in their content may have reduced support and positive perceptions, which could also apply to other interventions where the government is regulating platforms. For example, requiring platforms to release data and internal research was viewed as more intrusive than when platforms do so without a government requirement.

The analysis of individual differences in support for and perceptions of interventions revealed that men tend to support misinformation interventions less than women do, and that Republicans and Independents support them less compared to Democrats. In addition to supporting interventions less, men and Republicans view them as less fair, less effective, and more intrusive than do women and Democrats, respectively. The gender gap reflects broader trends of women supporting more (government) regulations than men across policy domains (Pew Research Center 2012). The gap in support is, in part, explained by differences in perceptions of the proposed interventions. Future work should examine how perceived features of policies interact with other factors like emotional reactions and issue awareness to predict differences in support between genders (Schlesinger and Heldman 2001).

The partisan differences align with previous findings (Appel, Pan, and Roberts 2023; Saltz et al. 2021; Newman et al. 2025). Previous studies have shown that Republicans are more likely to perceive interventions as biased against them (Saltz et al. 2021; Vogels 2022). A 2022 Pew Research poll found that approximately 70% of Republicans believe that major technology companies favor the views of liberals over conservatives, while only 22% of Democrats say they believe companies favor conservatives over liberals (Vogels 2022). While there is some evidence that Republican content and accounts are disproportionately removed, it may be because they tend to share lower quality news (as rated by politically balanced groups of laypeople) (Haimson et al. 2021; Mosleh et al. 2024). Therefore, it is likely that because Republicans believe they are more likely to be censored for their viewpoints, they perceive interventions as less fair and are less supportive of all interventions potentially employed by companies or the government. Whether this perception is accurate or not, it is imperative that social media companies work to (re)build trust among all their users when introducing interventions against misinformation. At the same time, although we observe several significant partisan interactions, the overall pattern of results does not differ across parties.

Fairness emerged as the most significant predictor of support for interventions, and as especially important for Republicans and men. This emphasis on fairness is not unique to misinformation and social media policies. Across a variety of policy contexts, including health, environment, and transportation, perceived fairness and effectiveness are often among the most predictive factors associated with support (Bamberg and Rölle 2003; Bergquist et al. 2022; Bos et al. 2015; Grelle and Hofmann 2024). In fact, a systematic review found that enhancing the communication of a policy's effectiveness to participants can boost support levels by 4%, a small but meaningful increase (Reynolds et al. 2020).

Intrusiveness is also a related factor, but it is not surprising that it holds less importance in the social media space than in other policy contexts. Previous studies have shown that when interventions are perceived to more directly impact an individual's personal choices or daily life, like increased costs associated with owning a car (Kallbekken, Garcia, and Korneliussen 2013), they tend to be unpopular (Diepeveen et al. 2013; Hagmann, Siegrist, and Hartmann 2018; Huber, Wicki, and Bernauer 2020). It may be that the potential intrusiveness of social media policies on the user experience results in less severe consequences for the typical individual than in other policy areas, which could impact access to food, health care, transportation, and more.

## 6 Limitations and Future Work

While this study is among the first to analyze the factors associated with support for misinformation interventions, several limitations could be addressed in future work. First, we focused our survey on active social media users, as those individuals would be the most affected by any potential policies and the most familiar with current interventions. Our sample was a convenience sample of US residents recruited via CloudResearch's MTurk Toolkit and was not designed to be demographically representative of US social media users. MTurk samples are known to differ from population benchmarks on characteristics such as age, education, and partisanship (e.g., Stagnaro et al. (2024)), and although our participants' demographics and platform use are broadly consistent with recent estimates for US social media users (Gottfried and Park 2025), generalizations should be made with caution.

Moreover, we only examined preferences in a single cultural and national context. The US is a relevant and important context to study public support for misinformation interventions. Many of the largest social media platforms (e.g., Facebook, X) are headquartered in the US, so US legal and political debates about content moderation (e.g., around Section 230) likely shape how platforms design and justify their policies globally. Furthermore, cross-national work on resilience to disinformation suggests that the US may be especially vulnerable due to high political polarization and a fragmented commercial media system (Humprecht, Esser, and Van Aelst 2020). However, there is reason to expect preferences over who should intervene against misinformation, and how, to vary across cultures and regions. For instance, Germans tend to attribute greater responsibility to the government to respond to problematic online content than Americans do (Riedl et al. 2021), plausibly increasing their support for interventions implemented by government rather than companies. We therefore view our findings as characterizing the contemporary US context and encourage future research to investigate how and why intervention preferences vary across regions.

We also only included ten interventions to limit the length of the survey, allowing us to focus more on the factors behind support. However, several new and emerging interventions were not included. For example, X's Community Notes program has been relatively successful at increasing the volume of fact-checks and boosting trust in misinformation flags by using crowd-sourced misinformation detection and labels (Chuai et al. 2024; Drolsbach, Solovev, and Pröllochs 2024), although reports of its overall effectiveness in reducing engagement with misleading content are mixed (Chuai et al. 2024; Kankham and Hou 2024). Future research should supplement these results by surveying a wider range of interventions. In addition, we only examined self-reported support and perceived effectiveness, which may not map directly onto how people respond to interventions in real-world contexts. Prior work suggests that even individuals who report distrusting fact-checkers can still be influenced by fact-checks (Martel and Rand 2024), underscoring this potential gap between attitudes and behavior. Future research should further investigate how perceived fairness, effectiveness, and intrusiveness relate to behavioral responses to specific interventions, not only to stated support.

Furthermore, the framing of and details included in intervention descriptions may have affected responses. Research shows that policies are more accepted when their goals or methods are clearly explained (Reynolds et al. 2020), which could be varied in future studies of public support. We also indicated that the implementer of the intervention is responsible for misinformation detection, which is an oversimplification. Future work should assess how people think misinformation should be detected, and how this influences their support for downstream interventions. Another limitation is that we did not directly assess participants' familiarity with the specific interventions we described. Therefore, some ratings, especially for more abstract or less visible interventions, may partly reflect uncertainty or ambiguity rather than well-formed preferences. This highlights the need for future work that explicitly measures participants' familiarity with intervention types and examines how such knowledge shapes their evaluations.

Finally, we focused exclusively on effectiveness, fairness, and intrusiveness in a correlational study. More factors (e.g., transparency) should be considered in future surveys about public acceptance of policies. Problem awareness and general support for government intervention may also be important factors to consider in future research (Diepeveen et al. 2013; Grelle and Hofmann 2024). Additionally, while the implementer of the intervention was manipulated between participants, the associated factors (effectiveness, fairness, and intrusiveness) were not. Although the extensive literature in the policy acceptance field suggests that these three factors are considered determinants of support (Huber, Wicki, and Bernauer 2020; Grelle and Hofmann 2024), we cannot conclusively rule out the possibility that support also influences perceptions of these factors. Future research should experimentally manipulate these factors and include relevant control variables to determine their actual causal effects.

## 7 Conclusion

Collectively, our findings suggest that fairness is likely valued above intrusiveness and effectiveness when considering support for misinformation interventions, and it is especially critical for specific groups like Republicans. When designing and implementing misinformation interventions, mitigating any possible disproportionate impacts on certain groups or individuals is critical. In addition, public messaging should emphasize why each intervention is needed and how it is being implemented fairly, in addition to providing recourse for users when necessary. Furthermore, there is more support for and positive perceptions of interventions deployed by social media companies rather than the government, which may reflect broader trends in institutional trust. Most likely, effective misinformation interventions require collaboration across institutions. However, broader support for company-implemented interventions can be leveraged in public communications and education.

Our analysis of support levels and perceived features of interventions highlights the importance of promoting user agency to garner widespread support. For example, platforms can allow users to engage with misinformation warnings and nudges behind interstitials rather than strictly and opaquely removing violating content. At the same time, interventions should be carefully designed and implemented to minimize disruption to the user experience. Overall, this work has important implications for designing misinformation interventions and messaging that will be positively received by social media users.

## Authors

**Catherine King** is a Postdoctoral Fellow in policy in the Software and Societal Systems Department at Carnegie Mellon University, where she also earned her PhD in Societal Computing (cking2@cs.cmu.edu).

**Samantha C. Phillips** is a Societal Computing PhD student in the Software and Societal Systems Department at Carnegie Mellon University (samanthp@cs.cmu.edu).

**Kathleen M. Carley** is a Professor of Computer Science in the Software and Societal Systems Department, IEEE Fellow, Director of the Center for Computational Analysis of Social and Organizational Systems (CASOS), and Director of the Center for Informed Democracy and Social-cybersecurity (IDeaS) at Carnegie Mellon University, and is the CEO of Netanomics.

## Acknowledgements

The authors thank the anonymous reviewers for their valuable feedback.

## Data availability statement

Replication files are available on Figshare.

- Data: https://figshare.com/s/aa7e43035af456300351
- Code: https://figshare.com/s/e31fd9581a070a893c1a
- Supplementary materials (including the survey questions): https://figshare.com/s/0edbc00f5e80becfc2ae

The study pre-analysis plan is registered at OSF: https://osf.io/b2yjt/

## Funding statement

This work was supported by the Knight Foundation; the Office of Naval Research's MURI: Persuasion, Identity, & Morality in Social-Cyber Environments grant N00014-21-12749; Carnegie Mellon University's Graduate Small Project Help (GuSH); the Center for Computational Analysis of Social and Organizational Systems (CASOS); and the Center for Informed Democracy and Social-cybersecurity (IDeaS). The views and conclusions contained in this document are those of the authors alone. The funders have no role in study design, data collection and analysis, decision to publish, or preparation of the manuscript.

## Ethical standards

This study, numbered "STUDY2022_00000143," was approved by the Carnegie Mellon University Institutional Review Board as exempt from a full review because no personally identifiable information was to be collected. Written informed consent was obtained from all participants.

# Appendices

## Appendix A: Regression model predicting intervention support (RQ1)

Table 3 reports the results from the regression model predicting support for a misinformation intervention as a function of perceived fairness, perceived effectiveness, perceived intrusiveness, and implementer (reference level: platform) with robust standard errors clustered on participant and intervention. We calculated adjusted fractional Bayes factors with Gaussian approximations for the primary models. We report $BF_{10}$ for each estimate where the alternative hypothesis is directional based on the sign of the estimate (i.e., $b < 0$, $b > 0$) and the null hypothesis is $b = 0$. Thus, if BF > 1, the evidence is more consistent with the alternative hypothesis; if BF < 1, the evidence is more consistent with the null hypothesis.

Table 3: OLS regression predicting support for a misinformation intervention, with robust standard errors clustered by participant and intervention (in parentheses), and the $BF_{10}$ for each estimate reported. *$p < 0.05$; **$p < 0.01$; ***$p < 0.001$.

| | Outcome: Support |
|---|---|
| Implementer (platform: government) | 0.123 (0.089) |
| | BF = 0.053 |
| Perceived fairness | 0.624 (0.016)*** |
| | BF > 100 |
| Perceived effectiveness | 0.302 (0.015)*** |
| | BF > 100 |
| Perceived intrusiveness | −0.065 (0.010)*** |
| | BF > 100 |
| Perceived fairness × implementer | −0.080 (0.024)*** |
| | BF = 5.91 |
| Perceived effectiveness × implementer | 0.077 (0.022)*** |
| | BF = 10.31 |
| Perceived intrusiveness × implementer | −0.036 (0.015)* |
| | BF = 0.42 |
| Intercept | 0.586 (0.060)*** |
| | BF > 100 |
| Observations | 8,071 |
| $R^2$ | 0.76 |
| Adjusted $R^2$ | 0.76 |

Table 4 on the following page displays the average marginal effects (AMEs) from the fixed effects model predicting support (Table 3). AMEs represent the average change in support associated with a one-unit increase in the predictor.

Table 4: Average marginal effects (AMEs) and 95% confidence intervals from the fixed effects model predicting support. *$p < 0.05$; **$p < 0.01$; ***$p < 0.001$.

| | AME (SE) | 95% CI |
|---|---|---|
| Implementer (platform: government) | −0.018 (0.015) | [−0.047, 0.011] |
| Perceived fairness | 0.584 (0.012)*** | [0.561, 0.608] |
| Perceived effectiveness | 0.341 (0.011)*** | [0.319, 0.362] |
| Perceived intrusiveness | −0.083 (0.007)*** | [−0.097, −0.068] |

**Robustness check:** Table 5 reports the results for our robustness check. This is the same model as Table 3 but includes responses from all participants who responded to at least part of the survey. The results are substantially similar to Table 3.

Table 5: Robustness check: OLS regression predicting support for a misinformation intervention, with robust standard errors clustered by participant and intervention (in parentheses), and the $BF_{10}$ for each estimate reported.
*$p < 0.05$; **$p < 0.01$; ***$p < 0.001$.

| | Outcome: Support |
|---|---|
| Implementer (platform: government) | 0.123 (0.089) |
| | BF = 0.054 |
| Perceived fairness | 0.626 (0.016)*** |
| | BF > 100 |
| Perceived effectiveness | 0.300 (0.015)*** |
| | BF > 100 |
| Perceived intrusiveness | −0.064 (0.010)*** |
| | BF > 100 |
| Perceived fairness × implementer | −0.083 (0.024)*** |
| | BF = 9.34 |
| Perceived effectiveness × implementer | 0.080 (0.022)*** |
| | BF = 17.93 |
| Perceived intrusiveness × implementer | −0.036 (0.015)* |
| | BF = 0.43 |
| Intercept | 0.582 (0.060)*** |
| | BF > 100 |
| Observations | 8,102 |
| $R^2$ | 0.761 |
| Adjusted $R^2$ | 0.76 |

Table 6 on the following page displays the average marginal effects (AMEs) from the fixed effects model predicting support (Table 5).

Table 6: Robustness check: Average marginal effects (AMEs) and 95% confidence intervals from the fixed effects model predicting support.
*$p < 0.05$; **$p < 0.01$; ***$p < 0.001$.

| | AME (SE) | 95% Confidence Interval |
|---|---|---|
| Implementer (platform: government) | −0.018 (0.015) | [−0.047, 0.011] |
| Perceived fairness | 0.585 (0.012)*** | [0.562, 0.608] |
| Perceived effectiveness | 0.340 (0.011)*** | [0.319, 0.362] |
| Perceived intrusiveness | −0.082 (0.007)*** | [−0.097, −0.068] |

## Appendix B: Comparing intervention support and perceptions (RQ2)

Figure 4 shows the support and perceived attribute ratings for each intervention broken up by party.

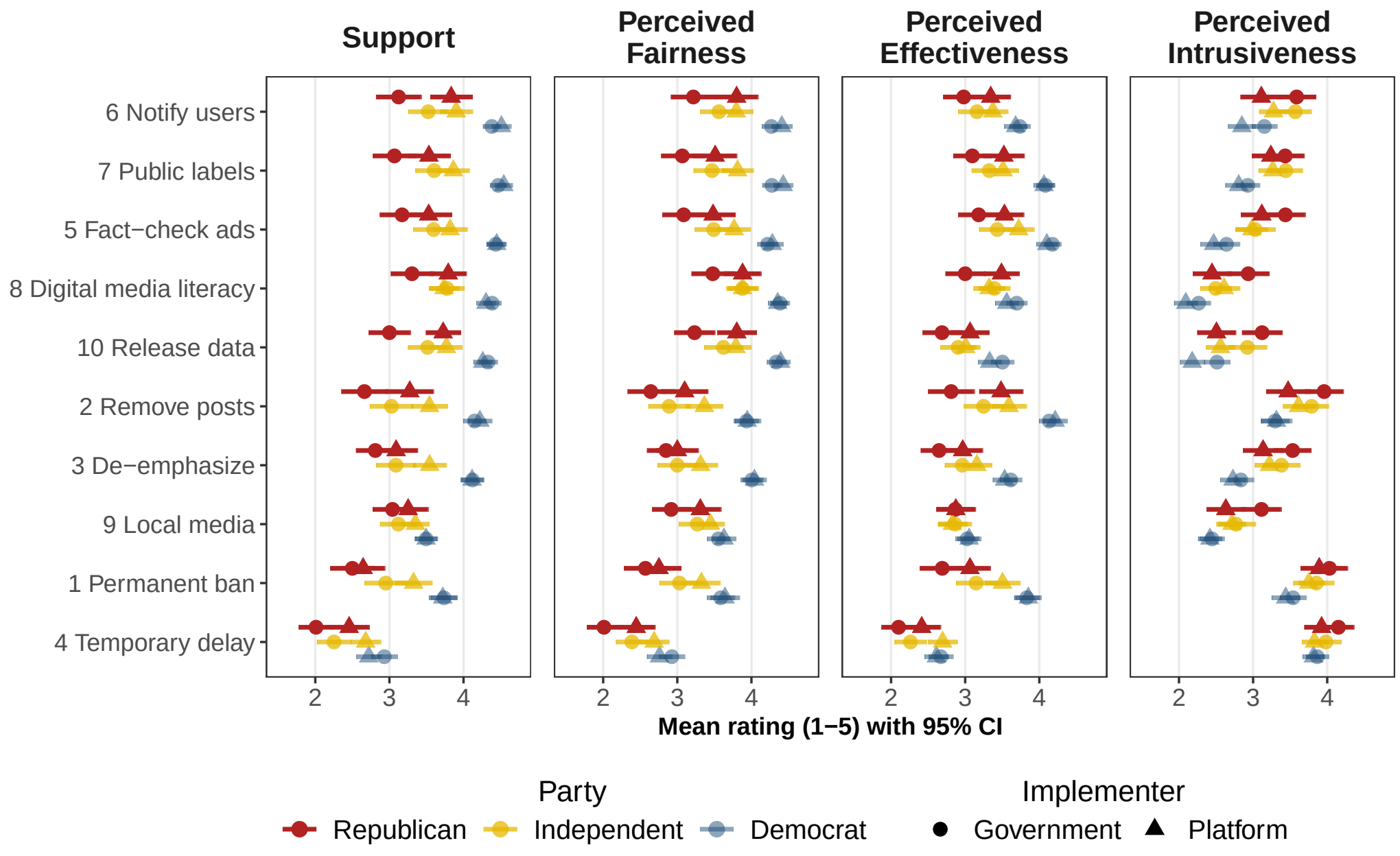

Figure 4: Average and 95% CI support, perceived fairness, perceived effectiveness, and perceived intrusiveness for each intervention (1–10) and implementer (government and platform) on a 1–5 Likert scale.

Table 7 reports the t-statistic from t-tests comparing average support, perceived fairness, perceived effectiveness, and perceived intrusiveness in platform vs. government implementation condition for each intervention. With the Bonferroni correction, 44 t-tests at $\alpha = 0.05$, $p < 0.0011$ is significant.

Table 7: T-statistic from t-tests comparing average rating values in platform vs. government implementation condition for each intervention.
*$p$ < 0.01; **$p$ < 0.001; ***$p$ < 0.0001.

| Item | Intervention Category | Support | Fairness | Effectiveness | Intrusiveness |
|---|---|---|---|---|---|
| 1 Permanent ban | Account Moderation | 1.22 | 1.49 | 2.01 | -1.13 |
| 2 Remove posts | Content Moderation | 2.52 | 1.84 | 2.46 | -1.33 |
| 3 De-emphasize | Content Moderation | 1.77 | 1.10 | 0.874 | -1.83 |
| 4 Temporary delay | Content Distribution | 1.39 | 1.23 | 2.06 | -1.64 |
| 5 Fact-check ads | Content Distribution | 1.49 | 2.11 | 1.37 | -1.70 |
| 6 Notify users | Content Labeling | 3.41** | 2.69* | 1.19 | -3.59** |
| 7 Public labels | Content Labeling | 2.12 | 2.79* | 1.35 | -1.46 |
| 8 Digital media literacy | Media Literacy | 0.31 | 0.596 | 0.141 | -1.60 |
| 9 Local media | External Structural Responses | 1.27 | 2.01 | -0.091 | -1.43 |
| 10 Release data | External Structural Responses | 2.20 | 2.17 | 0.202 | -4.4*** |
| Overall | | 5.48*** | 5.58*** | 3.63** | -6.08*** |

## Appendix C: Individual differences in support and perceptions (RQ3.1)

Table 8 shows the OLS regressions predicting average support for, perceived fairness of, perceived effectiveness of, and perceived intrusiveness of interventions as a function of age (18–24 to 65+ age brackets mapped to numeric values 0 to 5), gender (reference level: Female), partisanship (reference level: Democrat, combined "Other" party with "Independent"), political ideology (very liberal to very conservative categories mapped to numeric values 0 to 4), highest education level (Less than high school diploma to Doctorate or Professional Degree categories mapped to numeric values 0 to 6), income (less than \$20,000 to over \$200,000 income brackets mapped to numeric values 0 to 7), frequency seeing misinformation (never to very often categories mapped to numeric values 0 to 4), and platform usage (the number of platforms between 0 and 12 that are visited at least once a week).

These results are robust to different categorizations of political party. Table 9 displays the results of the same model but with "Independent" and "Other/unaffiliated" partisan categories separated, while Table 10 maps partisan categories to corresponding numeric values.

The results from one-way ANOVA tests comparing average support, perceived fairness, perceived effectiveness, and perceived intrusiveness across categories for each demographic variable measured categorically are shown in Table 11. Categories were collapsed such that there were at least 50 participants in each category, except the "Other" gender category, which had only 25 participants.

Fisher's exact test was used to assess the association between partisan identification and political ideology (Table 12).

Table 8: OLS regressions predicting average support, perceived fairness, perceived effectiveness, and perceived intrusiveness of interventions. The "Other" political party is combined with "Independents." N = 1010. *$p < 0.05$; **$p < 0.01$; ***$p < 0.001$.

| | Avg. support rating | Avg. fairness rating | Avg. effectiveness rating | Avg. intrusiveness rating |
|---|---|---|---|---|
| Age | 0.045* | 0.067** | 0.012 | 0.032 |
| | (0.021) | (0.022) | (0.020) | (0.019) |
| Gender (male : female) | −0.185*** | −0.124* | −0.195*** | 0.099* |
| | (0.053) | (0.054) | (0.049) | (0.048) |
| Gender (other : female) | −0.059 | −0.247 | −0.251 | 0.127 |
| | (0.172) | (0.175) | (0.158) | (0.154) |
| Education | 0.013 | 0.003 | 0.007 | 0.016 |
| | (0.022) | (0.022) | (0.020) | (0.019) |
| Income | 0.025 | 0.039* | 0.003 | −0.003 |
| | (0.016) | (0.016) | (0.015) | (0.014) |
| Party (Ind/other : Dem) | −0.244*** | −0.234** | −0.163* | 0.203** |
| | (0.074) | (0.075) | (0.068) | (0.066) |
| Party (Rep : Dem) | −0.275** | −0.284** | −0.194* | 0.159 |
| | (0.106) | (0.108) | (0.098) | (0.095) |
| Political ideology | −0.296*** | −0.264*** | −0.185*** | 0.130*** |
| | (0.036) | (0.037) | (0.033) | (0.033) |
| Misinformation exposure | −0.025 | −0.017 | −0.063* | 0.041 |
| | (0.027) | (0.027) | (0.024) | (0.024) |
| Weekly platform usage | 0.054*** | 0.049*** | 0.074*** | −0.012 |
| | (0.013) | (0.013) | (0.012) | (0.012) |
| (Intercept) | 3.924*** | 3.755*** | 3.570*** | 2.564*** |
| | (0.143) | (0.145) | (0.131) | (0.128) |
| $R^2$ | 0.257 | 0.217 | 0.183 | 0.096 |
| Adj. $R^2$ | 0.249 | 0.209 | 0.174 | 0.087 |
| Residual std. error (df=999) | 0.827 | 0.842 | 0.761 | 0.744 |
| F-statistic (df=10; 999) | 34.50*** | 27.63*** | 22.33*** | 10.64*** |

Table 9: OLS regressions predicting average support, perceived fairness, perceived effectiveness, and perceived intrusiveness of interventions. This specification separates the "Other" political party from "Independents." Standard errors in parentheses. N = 1010. *$p < 0.05$; **$p < 0.01$; ***$p < 0.001$.

| | Avg. support rating | Avg. fairness rating | Avg. effectiveness rating | Avg. intrusiveness rating |
|---|---|---|---|---|
| Age | 0.045* | 0.066** | 0.011 | 0.033 |
| | (0.021) | (0.022) | (0.020) | (0.019) |
| Gender (male : female) | −0.187*** | −0.126* | −0.197*** | 0.101* |
| | (0.043) | (0.054) | (0.049) | (0.048) |
| Gender (other : female) | −0.028 | −0.210 | −0.208 | 0.095 |
| | (0.174) | (0.177) | (0.160) | (0.155) |
| Education | 0.013 | 0.004 | 0.007 | 0.016 |
| | (0.022) | (0.022) | (0.020) | (0.019) |
| Income | 0.025 | 0.040* | 0.003 | −0.003 |
| | (0.016) | (0.016) | (0.015) | (0.014) |
| Party (Ind : Dem) | −0.225** | −0.212** | −0.137* | 0.184** |
| | (0.075) | (0.076) | (0.069) | (0.067) |
| Party (Other : Dem) | −0.485* | −0.514* | −0.495** | 0.443* |
| | (0.199) | (0.202) | (0.183) | (0.179) |
| Party (Rep : Dem) | −0.266* | −0.274* | −0.182 | 0.150 |
| | (0.106) | (0.108) | (0.098) | (0.095) |
| Political ideology | −0.299*** | −0.268*** | −0.190*** | 0.134*** |
| | (0.036) | (0.037) | (0.033) | (0.033) |
| Misinformation exposure | −0.026 | −0.018 | −0.063* | 0.041 |
| | (0.027) | (0.027) | (0.024) | (0.024) |
| Weekly platform usage | 0.053*** | 0.048*** | 0.072*** | −0.011 |
| | (0.013) | (0.013) | (0.012) | (0.012) |
| (Intercept) | 3.934*** | 3.766*** | 3.583*** | 2.555*** |
| | (0.143) | (0.145) | (0.131) | (0.128) |
| $R^2$ | 0.258 | 0.218 | 0.186 | 0.098 |
| Adj. $R^2$ | 0.250 | 0.210 | 0.177 | 0.088 |
| Residual std. error (df=998) | 0.827 | 0.841 | 0.760 | 0.744 |
| F-statistic (df=11; 998) | 31.54*** | 23.35*** | 20.70*** | 9.88*** |

Table 10: OLS regressions predicting average support, perceived fairness, perceived effectiveness, and perceived intrusiveness of interventions. Partisanship ranges from strong Democrat to strong Republican and is mapped to numeric values 0 to 4. Standard errors in parentheses. N = 1010. **p* < 0.05; ***p* < 0.01; ****p* < 0.001.

| | Avg. support rating | Avg. fairness rating | Avg. effectiveness rating | Avg. intrusiveness rating |
|---|---|---|---|---|
| Age | 0.045* | 0.067** | 0.011 | 0.032 |
| | (0.021) | (0.022) | (0.020) | (0.019) |
| Gender (male : female) | −0.183*** | −0.122* | −0.193*** | 0.099* |
| | (0.053) | (0.054) | (0.049) | (0.048) |
| Gender (other : female) | −0.065 | −0.248 | −0.252 | 0.140 |
| | (0.172) | (0.175) | (0.158) | (0.155) |
| Education | 0.013 | 0.002 | 0.006 | 0.015 |
| | (0.022) | (0.022) | (0.020) | (0.019) |
| Income | 0.024 | 0.039* | 0.002 | −0.003 |
| | (0.016) | (0.016) | (0.015) | (0.014) |
| Partisanship | −0.122*** | −0.129*** | −0.089** | 0.078** |
| | (0.033) | (0.033) | (0.030) | (0.030) |
| Political ideology | −0.265*** | −0.229*** | −0.160*** | 0.109** |
| | (0.037) | (0.038) | (0.034) | (0.034) |
| Misinformation exposure | −0.027 | −0.018 | −0.063* | 0.043 |
| | (0.026) | (0.027) | (0.024) | (0.024) |
| Weekly platform usage | 0.057*** | 0.051*** | 0.076*** | −0.015 |
| | (0.013) | (0.013) | (0.012) | (0.012) |
| (Intercept) | 3.925*** | 3.765*** | 3.575*** | 2.590*** |
| | (0.139) | (0.142) | (0.128) | (0.126) |
| $R^2$ | 0.258 | 0.220 | 0.185 | 0.094 |
| Adj. $R^2$ | 0.252 | 0.213 | 0.177 | 0.086 |
| Residual std. error (df=1000) | 0.826 | 0.840 | 0.760 | 0.745 |
| F-statistic (df=9; 1000) | 38.71*** | 31.34*** | 25.17*** | 11.53*** |

Table 11: F-statistics (degrees of freedom) from one-way ANOVA tests comparing support, perceived fairness, perceived effectiveness, and perceived intrusiveness across categories for each demographic variable. *$p < 0.05$; **$p < 0.01$; ***$p < 0.001$.

| | Categories | Average support rating | Average fairness rating | Average effectiveness rating | Average intrusiveness rating |
|---|---|---|---|---|---|
| Age | 18–34<br>35–44<br>45–54<br>55–64<br>65+ | F(4,1005) = 1.006 | F(4,1005) = 1.268 | F(4,1005) = 1.686 | F(4,1005) = 2.707* |
| Gender | Men<br>Women<br>Other | F(2,1007) = 8.980*** | F(2,1007) = 4.992** | F(2,1007) = 10.282*** | F(2,1007) = 2.760 |
| Education | High school or less<br>Some college<br>Associate’s degree<br>Bachelor’s degree<br>Master’s degree or higher | F(4,1005) = 3.027* | F(4,1005) = 2.094 | F(4,1005) = 2.029 | F(4,1005) = 0.205 |
| Income | Less than $20,000<br>$20,000–$39,999<br>$40,000–$59,999<br>$60,000–$79,999<br>$80,000–$99,999<br>$100,000–$149,999<br>Over $150,000 | F(6,1003) = 1.400 | F(6,1003) = 1.636 | F(6,1003) = 0.806 | F(6,1003) = 1.132 |
| Partisanship | Strong Democrat<br>Weak Democrat<br>Independent/other<br>Weak Republican<br>Strong Republican | F(4,1005) = 61.972*** | F(4,1005) = 51.508*** | F(4,1005) = 34.935*** | F(4,1005) = 21.284*** |
| Political ideology | Very liberal<br>Liberal<br>Moderate<br>Conservative<br>Very conservative | F(4,1005) = 71.889*** | F(4,1005) = 58.054*** | F(4,1005) = 36.968*** | F(4,1005) = 21.980*** |
| Misinformation exposure | Never<br>Rarely<br>Sometimes<br>Often<br>Very often | F(4,1005) = 0.177 | F(4,1005) = 0.360 | F(4,1005) = 0.469 | F(4,1005) = 0.479 |
| Weekly platform usage | Q1: 0–3 platforms<br>Q2: 4–5 platforms<br>Q3: 6 platforms<br>Q4: 7–11 platforms | F(3,1006) = 8.896*** | F(3,1006) = 6.682*** | F(3,1006) = 16.047*** | F(3,1006) = 6.455*** |

Table 12: Contingency table of partisan affiliation and political ideology groups. Fisher's exact test comparing party and ideology categories yields $\chi^2 = 107.2$, $p < 0.001$.

| | Republican | Independent/other | Democrat |
|---|---|---|---|
| Very liberal | 4 | 17 | 148 |
| Liberal | 3 | 37 | 265 |
| Moderate | 20 | 206 | 46 |
| Conservative | 138 | 40 | 9 |
| Very conservative | 71 | 3 | 3 |

The average ratings for the remaining demographic variables are shown in Figure 5 on the following page.

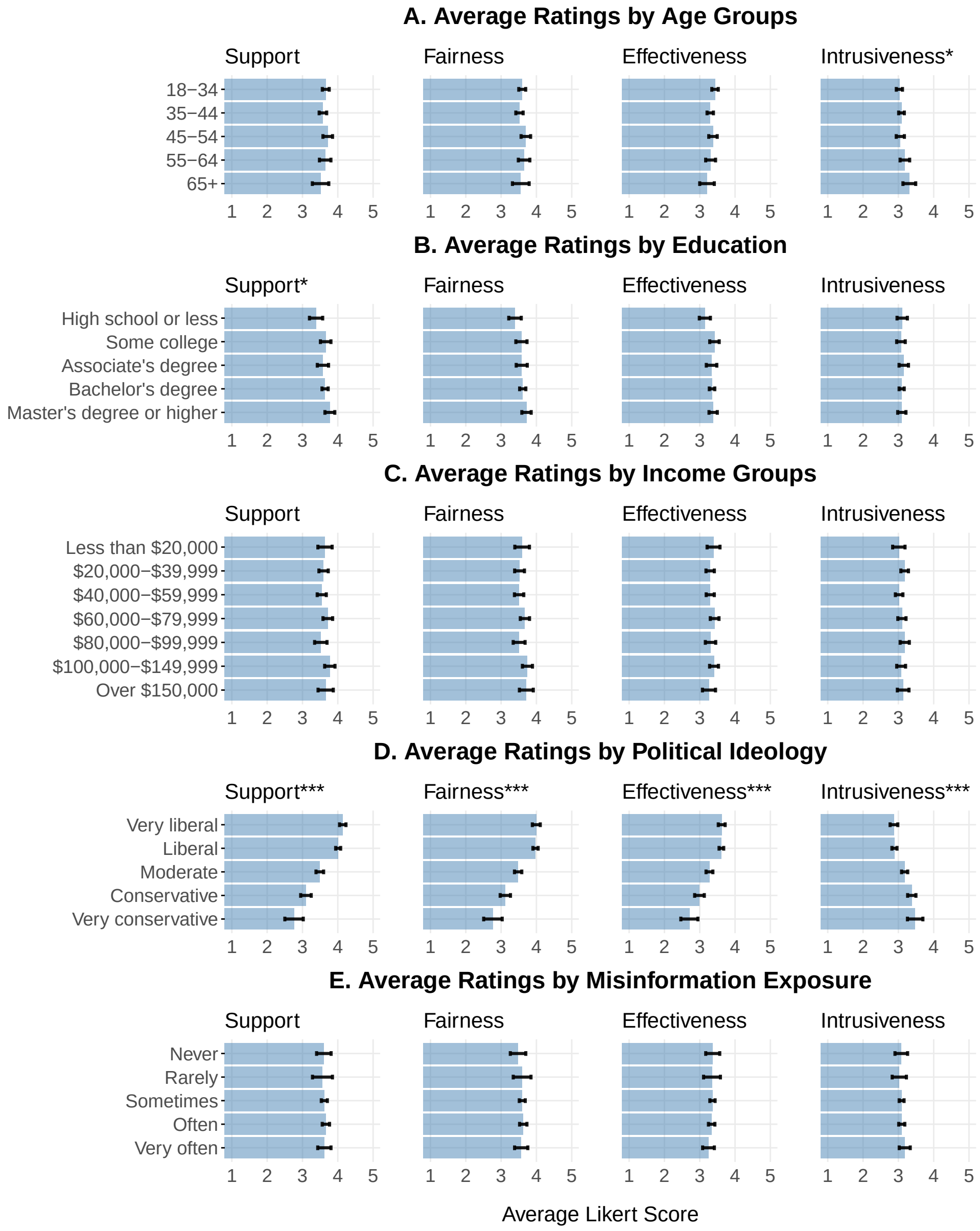

Figure 5: Average ratings (95% CI) across various demographic groups. One-way ANOVA tests were run for each variable and outcome, with stars indicating the level of significance: *$p < 0.05$; **$p < 0.01$; ***$p < 0.001$.

## Appendix D: Individual differences in attributes associated with support (RQ3.2)

Table 13 shows the results from the OLS regression predicting support for a misinformation intervention as a function of perceived fairness, perceived effectiveness, perceived intrusiveness, implementer (reference level: social media company), gender (reference level: Female), partisanship (reference level: Democrat, with Independents combined with "Other") and platform usage (the number of platforms between 0 and 12 that are visited at least once a week). Table 14 shows the results from the same model, except with political ideology instead of partisanship.

Table 13: OLS regression predicting support for a misinformation intervention with robust standard errors clustered on participant and intervention. The "Other" political party is combined with "Independents." N = 1010. *$p < 0.05$; **$p < 0.01$; ***$p < 0.001$.

| | Outcome: Support<br>Estimate (standard error) |
|---|---|
| **Main effects** | |
| Implementer (government : platform) | 0.193 (0.098)* |
| Perceived fairness | 0.503 (0.037)*** |
| Perceived effectiveness | 0.363 (0.035)*** |
| Perceived intrusiveness | −0.111 (0.024)*** |
| Gender (male : female) | −0.232 (0.089)** |
| Gender (other : female) | 0.653 (0.293)* |
| Party (Independent/other : Democrat) | −0.288 (0.117)* |
| Party (Republican : Democrat) | −0.255 (0.112)* |
| Weekly platform usage | −0.034 (0.021) |
| **Interactions with implementer** | |
| Implementer × perceived fairness | −0.087 (0.024)*** |
| Implementer × perceived effectiveness | 0.063 (0.022)** |
| Implementer × perceived intrusiveness | −0.035 (0.014)* |
| Implementer × gender (male : female) | 0.009 (0.030) |
| Implementer × gender (other : female) | 0.186 (0.107) |
| Implementer × party (Independent/other : Democrat) | −0.103 (0.037)** |
| Implementer × party (Republican : Democrat) | −0.094 (0.039)* |
| Implementer × weekly platform usage | 0.006 (0.007) |
| **Interactions with perceived fairness** | |
| Fairness × gender (male : female) | 0.048 (0.024)* |
| Fairness × gender (other : female) | −0.068 (0.060) |
| Fairness × party (Independent/other : Democrat) | 0.047 (0.030) |
| Fairness × party (Republican : Democrat) | 0.065 (0.029)* |
| Fairness × weekly platform usage | 0.011 (0.005)* |
| **Interactions with perceived effectiveness** | |
| Effectiveness × gender (male : female) | −0.016 (0.022) |
| Effectiveness × gender (other : female) | −0.051 (0.068) |
| Effectiveness × party (Independent/other : Democrat) | 0.011 (0.027) |
| Effectiveness × party (Republican : Democrat) | −0.024 (0.027) |
| Effectiveness × weekly platform usage | −0.011 (0.005)* |
| **Interactions with perceived intrusiveness** | |
| Intrusiveness × gender (male : female) | 0.021 (0.015) |
| Intrusiveness × gender (other : female) | −0.048 (0.047) |
| Intrusiveness × party (Independent/other : Democrat) | −0.009 (0.019) |
| Intrusiveness × party (Republican : Democrat) | −0.018 (0.018) |
| Intrusiveness × weekly platform usage | 0.009 (0.003)** |
| (Intercept) | 1.075 (0.154)*** |
| Observations | 8,071 |
| RMSE | 0.656 |
| Adjusted $R^2$ | 0.766 |

Table 14: OLS regression predicting support for a misinformation intervention with robust standard errors clustered on participant and intervention. Political ideology is included instead of political party, with very liberal to very conservative categories mapped to numeric values 0 to 4. N = 1010. $^{*}p < 0.05$; $^{**}p < 0.01$; $^{***}p < 0.001$.

| | Outcome: Support<br>Estimate (standard error) |
|---|---|
| **Main effects** | |
| Implementer (government : platform) | 0.132 (0.103) |
| Perceived fairness | 0.482 (0.039)*** |
| Perceived effectiveness | 0.386 (0.036)*** |
| Perceived intrusiveness | −0.117 (0.025)*** |
| Gender (male : female) | −0.218 (0.088)* |
| Gender (other : female) | 0.576 (0.286)* |
| Ideology | −0.131 (0.038)*** |
| Weekly platform usage | −0.031 (0.020) |
| **Interactions with implementer** | |
| Implementer × perceived fairness | −0.077 (0.024)** |
| Implementer × perceived effectiveness | 0.059 (0.022)** |
| Implementer × perceived intrusiveness | −0.035 (0.014)* |
| Implementer × gender (male : female) | 0.001 (0.030) |
| Implementer × gender (other : female) | 0.172 (0.104) |
| Implementer × ideology | −0.017 (0.014) |
| Implementer × weekly platform usage | 0.009 (0.007) |
| **Interactions with perceived fairness** | |
| Fairness × gender (male : female) | 0.041 (0.023) |
| Fairness × gender (other : female) | −0.066 (0.060) |
| Fairness × ideology | 0.030 (0.010)** |
| Fairness × weekly platform usage | 0.011 (0.005)* |
| **Interactions with perceived effectiveness** | |
| Effectiveness × gender (male : female) | −0.009 (0.022) |
| Effectiveness × gender (other : female) | −0.049 (0.067) |
| Effectiveness × ideology | −0.017 (0.009)* |
| Effectiveness × weekly platform usage | −0.010 (0.005)* |
| **Interactions with perceived intrusiveness** | |
| Intrusiveness × gender (male : female) | 0.019 (0.014) |
| Intrusiveness × gender (other : female) | −0.038 (0.047) |
| Intrusiveness × ideology | 0.002 (0.006) |
| Intrusiveness × weekly platform usage | 0.008 (0.003)** |
| (Intercept) | 1.156 (0.159)*** |
| Observations | 8,071 |
| RMSE | 0.654 |
| Adjusted $R^2$ | 0.767 |

## Appendix E: Sample statistics

Table 15: Sample composition of participants who completed the survey (N = 1,010). For categorical variables, entries report number (percentage). The total percentage may not sum to exactly 100% due to rounding. For numerical variables, entries report mean (standard deviation).

| Variable | N (%) or Mean (SD) |
|---|---|
| **Gender** | |
| Female | 465 (46.0%) |
| Male | 520 (51.5%) |
| Transgender male | 7 (0.7%) |
| Transgender female | 2 (0.2%) |
| Non-binary / third gender | 7 (0.7%) |
| Other | 1 (0.1%) |
| Prefer not to respond | 8 (0.8%) |
| **Age** | |
| 18–24 | 28 (2.8%) |
| 25–34 | 225 (22.2%) |
| 35–44 | 338 (33.5%) |
| 45–54 | 186 (18.4%) |
| 55–64 | 148 (14.7%) |
| 65+ | 85 (8.4%) |
| **Hispanic, Latino, or Spanish origin** | |
| Yes | 77 (7.6%) |
| No | 933 (92.4%) |
| **Race (check all that apply)** | |
| White / Caucasian | 838 (83.0%) |
| Black or African American | 98 (9.7%) |
| Asian | 83 (8.2%) |
| American Indian or Alaska Native | 26 (2.6%) |
| Native Hawaiian or Pacific Islander | 2 (0.2%) |
| Other | 13 (1.3%) |
| **Highest level of education** | |
| Less than high school diploma | 4 (0.4%) |
| High school diploma or equivalent | 112 (11.1%) |

*Continued on next page*

| Variable | N (%) or Mean (SD) |
|---|---|
| Some college, no degree | 164 (16.2%) |
| Associate degree | 126 (12.5%) |
| Bachelor's degree | 438 (43.4%) |
| Master's degree | 134 (13.3%) |
| Doctorate | 15 (1.5%) |
| Professional degree (e.g., MD, DDS) | 17 (1.7%) |
| **Household income before taxes** | |
| Less than $20,000 | 87 (8.6%) |
| $20,000–39,999 | 172 (17.0%) |
| $40,000–59,999 | 220 (21.8%) |
| $60,000–79,999 | 181 (17.9%) |
| $80,000–99,999 | 140 (13.9%) |
| $100,000–149,999 | 131 (13.0%) |
| $150,000–199,999 | 48 (4.8%) |
| Over $200,000 | 31 (3.1%) |
| **Current religion** | |
| Protestant | 270 (26.7%) |
| Roman Catholic | 170 (16.8%) |
| Mormon | 6 (0.6%) |
| Orthodox Christian | 7 (0.7%) |
| Jewish | 22 (2.2%) |
| Muslim | 11 (1.1%) |
| Buddhist | 16 (1.6%) |
| Hindu | 5 (0.5%) |
| Atheist | 154 (15.2%) |
| Agnostic | 187 (18.5%) |
| Nothing in particular | 119 (11.8%) |
| Other | 43 (4.3%) |
| **Political party affiliation** | |
| Democrat | 471 (46.6%) |
| Republican | 236 (23.4%) |
| Independent | 284 (28.1%) |
| Other | 19 (1.9%) |

*Continued on next page*

| Variable | N (%) or Mean (SD) |
|---|---|
| **Strength of party affiliation (Democrats/Republicans only)** | |
| Strong | 398 (39.4%) |
| Not strong | 309 (30.5%) |
| **Lean of Independents** | |
| Closer to Republican Party | 128 (12.7%) |
| Closer to Democratic Party | 171 (16.9%) |
| **Political ideology** | |
| Very liberal | 169 (16.7%) |
| Liberal | 305 (30.2%) |
| Moderate | 272 (26.9%) |
| Conservative | 187 (18.5%) |
| Very conservative | 77 (7.6%) |
| **Platforms used at least weekly** | |
| Facebook | 758 (75.0%) |
| X (formerly Twitter) | 544 (53.9%) |
| YouTube | 893 (88.4%) |
| Pinterest | 206 (20.4%) |
| Instagram | 645 (63.9%) |
| Reddit | 606 (60.0%) |
| LinkedIn | 238 (23.6%) |
| TikTok | 354 (35.0%) |
| Snapchat | 162 (16.0%) |
| WhatsApp | 179 (17.7%) |
| Nextdoor | 137 (13.6%) |
| Other | 24 (2.4%) |
| **Platform usage (means)** | |
| Number of platforms visited (total) | 7.19 (2.21) |
| Visited at least monthly | 5.76 (2.22) |
| Visited at least weekly | 5.06 (2.15) |
| Visited at least daily | 3.45 (1.91) |